\documentclass[english,aps,prl,preprint,superscriptaddress]{revtex4-1}
\usepackage[T1]{fontenc}
\usepackage[latin9]{inputenc}
\usepackage{amsmath}
\usepackage{amssymb}
\usepackage{graphicx}
\usepackage{wasysym}
\usepackage{subscript}

\makeatletter

\providecommand{\tabularnewline}{\\}

\usepackage{pslatex}

\usepackage{babel}

\usepackage{babel}

\makeatother

\usepackage{babel}
\begin{document}
\title{Determination of CaOH and CaOCH\textsubscript{3} vibrational branching
ratios for direct laser cooling and trapping}
\author{Ivan Kozyryev}
\email{ivan@cua.harvard.edu}

\altaffiliation{Present address: Department of Physics, Columbia University, New York, NY 10027}

\affiliation{Harvard-MIT Center for Ultracold Atoms, Cambridge, MA 02138}
\affiliation{Department of Physics, Harvard University, Cambridge, MA 02138}
\author{Timothy C. Steimle}
\email{tsteimle@asu.edu}

\affiliation{School of Molecular Sciences, Arizona State University, Tempe, AZ
85287}
\author{Phelan Yu}
\affiliation{Harvard-MIT Center for Ultracold Atoms, Cambridge, MA 02138}
\affiliation{Department of Physics, Harvard University, Cambridge, MA 02138}
\author{Duc-Trung Nguyen }
\affiliation{School of Molecular Sciences, Arizona State University, Tempe, AZ
85287}
\author{John M. Doyle}
\affiliation{Harvard-MIT Center for Ultracold Atoms, Cambridge, MA 02138}
\affiliation{Department of Physics, Harvard University, Cambridge, MA 02138}
\date{\today}
\begin{abstract}
Alkaline earth monoalkoxide free radicals (MORs) have molecular properties
conducive to direct laser cooling to sub-millikelvin temperatures.
Using dispersed laser induced fluorescence (DLIF) measurements from
a pulsed supersonic molecular beam source we determine vibrational
branching ratios and Franck-Condon factors for the MORs CaOH and CaOCH\textsubscript{3}.
With narrow linewidth continuous-wave dye laser excitation, we precisely
measure fluorescence branching for both $\tilde{X}-\tilde{A}$ and
$\tilde{X}-\tilde{B}$ electronic systems in each molecule. Weak symmetry-forbidden
decays to excited bending states with non-zero vibrational angular
momentum are observed. Normal mode theoretical analysis combined with
\textit{ab initio} structural calculations are performed and compared
to experimental results. Our measurements and analysis pave the way
for direct laser cooling of these (and other) complex nonlinear polyatomic
molecules. We also describe a possible approach to laser cooling and
trapping of molecules with fewer symmetries like chiral species. 
\end{abstract}
\maketitle

\section{introduction }

Detailed quantum mechanical understanding of dynamics, interactions,
and reactions for complex molecules requires precisely targeted preparation
of internal molecular states and relative kinetic motions in the gas-phase
solvent-free environment \cite{Krems2008,balakrishnan2016perspective,bohn2017cold}.
Supersonic molecular beams have allowed groundbreaking achievements
in experimental studies of inelastic and reactive molecular interactions,
as well as opened the path to coherent control of internal molecular
states \cite{Narevicius2012,narevicius2014observation,narevicius2015molecular,narevicius2017directly,zare2018supersonic,zare2017quantum}.
With some crucial exceptions, however, large velocity in the laboratory
frame usually leads to short experimental coherence times and averaging
over multiple internal and motional quantum states. The emergence
of cryogenic buffer-gas beams (CBGBs) led to a significant reduction
in the forward velocity but the highest molecular fluxes still maintained
$\apprge10$ K kinetic energy in the laboratory frame \cite{Hutzler2012CBGB}.
Therefore, other experimental techniques have emerged in particular
those aimed at reducing the forward velocity of molecular beams to
the degree sufficient for eventual three-dimensional confinement inside
conservative electromagnetic traps, enabling seconds-long coherence
times \cite{Lemeshko2013review}. Molecular trapping also opens opportunities
for accurate collisional studies at very low temperatures.

The use of electric, magnetic, optical and mechanical techniques for
molecular beam slowing have all been experimentally demonstrated,
with published comprehensive reviews providing excellent survey of
the field \cite{jankunas2015cold,Lemeshko2013review}. Recent advances
in direct molecular cooling as well as coherent molecule formation
from unbound atoms have led to the emergence and rapid development
of a new vibrant field of cold and ultracold chemistry \cite{bohn2017cold,jin2011polar}.
The unprecedented opportunity to completely control the internal molecular
degrees of freedom, as well as external motion, of diverse molecular
species has yielded exciting observations on the role of quantum statistics
in molecular reactive collisions as well as confining geometry of
the reactants, among other results \cite{bohn2017cold}. However,
with few crucial exceptions \cite{zare2017quantum,narevicius2015molecular,campbell2009mechanism,egorov2005zeeman},
the studies have been limited to the exploration of diatomic molecules
which can be prepared from associated laser-cooled atomic samples.

Extension of the available experimental techniques to production of
a chemically diverse set of polyatomic molecules will aid in benchmarking
state-of-the-art\textit{ ab initio} calculations as well as shed light
on novel reaction mechanisms at work. Given the recent rapid experimental
progress on laser cooling and trapping of diatomic molecules \cite{Shuman2010SrF,Hummon2013YO,truppe2017molecules,anderegg2017radio,anderegg2018laser},
we consider the extension of these techniques to polyatomic species.
The internal quantum complexity of polyatomic molecules grows significantly
upon transitioning from polar diatomic radicals with a single vibrational
mode to nonlinear polyatomic molecules with multiple rotational and
vibrational degrees of freedom. Nevertheless, a large number of different
monovalent alkaline earth derivatives have been proposed to be amenable
to laser cooling~\cite{Kozyryev2016MOR,isaev2015polyatomic}, with
pioneering experimental work on the laser-cooling of the triatomic
SrOH molecule \cite{Kozyryev2017}.

In polyatomic molecules, the absence of strict angular momentum selection
rules for transitions between totally symmetric vibrational modes
presents a new challenge for photon cycling \cite{herzberg1966molecular},
a necessary ingredient for laser cooling. Additionally, higher-order
perturbation mechanisms present only in polyatomic molecules such
as Fermi resonance as well as Jahn-Teller and Renner-Teller interactions
can lead to Born-Oppenheimer breakdown and coupling between vibrational
modes, which can result in loss channels otherwise forbidden by the
symmetries of the original unperturbed states. Thus, precise measurements
of Franck-Condon factors (FCFs) and vibrational branching ratios (VBRs)
are crucial in determining the feasibility of laser cooling.

While there is plenty of theoretical work on the estimation of Franck-Condon
factors and vibrational branching ratios for both diatomic \cite{lane2011electronic,lane2018quantitative,gao2014laser}
and polyatomic molecules \cite{isaev2015polyatomic,isaev2016laser,Kozyryev2016MOR},
accurate experimental studies have been limited primarily to diatomic
species \cite{tarbutt2014BH,tarbutt2014radiative,zhuang2011franck,hunter2012prospects}
and, recently, SrOH \cite{nguyen2018fluorescence}. Accurate theoretical
predictions of FCFs for polyatomic molecules remain a challenge because
their multidimensional nature not only introduces interactions between
the degrees of freedom on each potential surface but also between
the coordinates of the two electronic states. The pair of molecules
we have chosen to study here represent the simplest (by geometric
structure) members of the alkaline earth monoalkoxide radicals (MOR)
family \cite{brazier1985laser,brazier1986laser}, a large class of
polyatomic molecules that has previously been proposed for laser cooling
applications \cite{Kozyryev2016MOR}. A simple ``triatomic'' model
of the MOR molecules has been used in order to estimate branching
ratios and indicated that laser cooling of relatively large MOR molecules
with up to 15 constituent atoms could be possible. Our precise measurements
of vibrational branching ratios for CaOH and CaOCH\textsubscript{3}
described below provide the first experimental validation of the theoretical
model of ``shielding'' of the R ligand group (e.g. CH\textsubscript{3})
vibrational modes by the intermediate oxygen atom \cite{Kozyryev2016MOR}.
Increased density of vibrational states for larger polyatomic molecules
could potentially lead to internal vibrational redistribution among
multiple vibrational normal modes leading to enhanced loss probabilities
and, therefore, necessitating accurate experimental studies.

CaOH is one of the first polyatomic molecules to be proposed suitable
for laser cooling and trapping and detailed theoretical ab initio
calculations have been performed of the Franck-Condon factors \cite{isaev2015polyatomic}.
Our work provides an important comparison between the theoretical
and experimental results, benchmarking theoretical calculations by
Isaev and Berger \cite{isaev2015polyatomic}. Our precise measurements
of the branching ratios for a more complex symmetric-top molecule
(STM) CaOCH\textsubscript{3} provide, to the best of our knowledge,
the first accurate experimental study of the vibrational branching
ratios for a nonlinear radical suitable for direct laser cooling.
Particularly, our results indicate the necessity to consider pseudo
Jahn-Teller couplings in the excited electronic levels of polyatomic
molecules \cite{fischer1984vibronic}. Perhaps surprisingly, despite
the presence of Jahn-Teller interaction in the electronically degenerate
$E$ symmetry state of the $C_{3v}$ symmetric top, using the lowest
allowed $\tilde{X}^{2}A_{1}\leftrightarrow\tilde{A}^{2}E_{1/2}$ electronic
excitation will lead to $\sim16$ scattered photons without any additional
repumping lasers. Based on our experimental measurements and theoretical
analysis, we propose a feasible route for producing large ultracold
samples of both radicals via laser cooling on the lower spin-orbit
branch of the $\tilde{X}-\tilde{A}$ electronic transition. Performing
a detailed study of inelastic and reactive collisions on increasingly
complex members of the CaOR molecule family would provide a unique
window into the scattering properties of fundamental ligand groups
in organic chemistry. 

\section{experimental configuration}

Since the details of the experimental apparatus have previously been
described in previous publications \cite{nguyen2018fluorescence},
we provide only a brief account of the experimental configuration
employed. As a source of cold molecules we used a pulsed supersonic
molecular beam with argon as a carrier gas. Both CaOH and CaOCH\textsubscript{3}
beams were created by flowing argon over a container filled with liquid
methanol (CH\textsubscript{3}OH) maintained at room temperature.
Room temperature vapor pressure of methanol ($\sim10$ kPa) together
with the backing pressure of argon ($\sim4,000$ kPa) was sufficient
to seed a sufficient number of gas-phase molecules into the carrier
gas to observe large, stable yields of both molecular radicals of
interest following laser ablation of a calcium metal target. The peak
fluorescence signal per single rovibrational level was approximately
of similar magnitude for CaOH and CaOCH\textsubscript{3}, which allowed
for accurate studies of both radicals using the same beam source.
Molecules in a specific rotational state were excited using a single
frequency cw-dye laser. Dispersed fluorescence data was collected
with a spectrometer that included 0.67-meter focal length, high-efficiency
Czerny-Turner-type monochromator with a low-dispersion grating and
a cooled, gated intensified CCD (ICCD) camera. The relative sensitivity
as a function of wavelength for the spectrometer was precisely calibrated
beforehand. The wavelength resolution of the emission spectrum was
controlled by adjusting the width of the monochromator input aperture
and absolute calibration was performed with the argon emission lamp.
The measured emission wavelengths agreed to $\lesssim0.1$ nm with
theoretical predictions which allowed for unambiguous identification
of the vibrational loss channels. The narrow linewidth ($\lesssim1$
MHz) of the excitation dye laser allowed accurate rotationally-resolved
excitation. Burleigh high-resolution wavemeter was used as a rough
frequency reference, additionally confirmed by co-recording with some
Doppler I\textsubscript{2} spectrum \cite{salumbides2006hyperfine}.
For each DLIF measurement, three separate datasets were taken: i)
signal dataset with with ablation Nd:YAG laser on and dye laser on,
ii) scattered background dataset with no Nd:YAG laser but dye laser
on, and iii) spurious ablation glow dataset with Nd:YAG laser present
but no dye laser. The resulting measurement was obtained by subtracting
datasets ii) and iii) from i) in order to remove the scattered light
offset and reduce the background ablation glow (e.g. metastable calcium
emission).

\section{experimental determination of branching ratios}

In order to eliminate rotational branching during the photon cycling
process the use of $J''\rightarrow J'-1$ type angular momentum transitions
has been proposed by Stuhl et al. \cite{stuhl2008magneto} and for
the first time demonstrated with diatomic SrF in Ref. \cite{shuman2009radiative}
and linear triatomic SrOH in Ref. \cite{kozyryev2016radiation}. Rotationally-resolved
high-resolution spectroscopy of low-$J$ rotational levels of CaOH
and CaOCH\textsubscript{3} has been previously performed and assigned
for both $\tilde{X}-\tilde{A}$ and $\tilde{X}-\tilde{B}$ electronic
bands \cite{bernath1984dye,crozet2002CaOCH3,crozet2005geometry,whitham1998laser,bernath1985CaOH,steimle1992supersonic,steimle1998spectroscopic}.
The availability of prior spectroscopic measurements, low rotational
temperatures of the skimmed supersonic molecular beam as well as the
narrow linewidth of the cw dye laser, allowed us to deliberately address
only the rotational transitions which can be used for optical cycling
and laser cooling. Additionally, the absence of state-changing collisions
in the probing region eliminated any possible systematic errors.

While both CaOH and CaOCH\textsubscript{3} have spin-rotation and
hyperfine splittings arising from the unpaired electron and hydrogen
spins, correspondingly, for the purposes of the experiments performed
here we label the states using rotational quantum numbers $\left|N,K\right\rangle $
for non-degenerate quantum states (i.e. $\Sigma^{+}$ states for linear
molecules and $A_{1}$ states for STMs) and $\left|J,K\right\rangle $
states for degenerate states (i.e. $\Pi$ electronic states for linear
molecules and $E$ states for STMs) with $K=0$ for linear molecules.
Upon the electronic excitation to a specific rotational level, the
spontaneous emission rate for a dipole-allowed transition is governed
by the Einstein A coefficient \cite{bernath2005spectra}: 
\begin{equation}
A_{J'\rightarrow J''}=\frac{16\pi^{3}\nu^{3}S_{J'J''}}{3\epsilon_{0}hc^{3}\left(2J'+1\right)}
\end{equation}
where the molecular line strength $S_{J'J''}\equiv\sum_{M',M''}\left|\left\langle J'M'\left|\mathbf{\mu}\right|J''M''\right\rangle \right|^{2}$
is approximately given as 
\begin{equation}
S_{J'J''}=q_{\mathbf{v'-v''}}\left|\mathbf{R}_{e}\right|^{2}S_{J''}^{\triangle J}
\end{equation}
under the Born-Oppenheimer separation $\psi_{{\rm tot}}=\psi_{{\rm el}}\psi_{{\rm vib}}\psi_{{\rm rot}}$.
Therefore, the intensity of different vibrational bands will be proportional
to 
\begin{equation}
VBR=\frac{\nu^{3}q_{\mathbf{v'-v''}}}{\sum_{i}\nu_{i}^{3}q_{\mathbf{v'-v''}}}
\end{equation}
since the same electronic transition dipole moment $\left|\mathbf{R}_{e}\right|$
and H�nl-London rotational factor are shared by all the vibronic emission
bands. Additionally, the rotational branching is limited by the $\triangle J=0,\pm1$
selection rule and all $J$ branches contribute identically for all
the vibrational bands. For polyatomic molecules, the Franck-Condon
factor is multidimensional and given as 
\begin{equation}
q_{\mathbf{v'-v''}}=q_{v_{1}'-v_{1}''}q_{v_{2}'-v_{2}''}\ldots=\left|\int\psi_{v_{1}'}^{*}\psi_{v_{1}''}dQ_{1}\right|^{2}\left|\int\psi_{v_{2}'}^{*}\psi_{v_{2}''}dQ_{2}\right|^{2}\ldots
\end{equation}
By comparing the integrated areas under the dispersed fluorescence
emission peaks, we can determine the relative branching ratios for
decays from a given electronic, vibrational and rotational state.

\subsection{CaOH measurements}

Two electronic transitions have been studies for CaOH that have potential
to be used in the laser cooling and trapping applications. Previously,
laser cooling of the isoelectronic CaF and SrOH molecules has been
demonstrated using either the $\tilde{X}^{2}\Sigma^{+}-\tilde{A}^{2}\Pi_{1/2}$
or $\tilde{X}^{2}\Sigma^{+}-\tilde{B}^{2}\Sigma^{+}$ electronic transition
and, therefore, we have performed detailed measurements on both bands
for CaOH as well. Additionally, in order to increase the photon cycling
rate it is favorable to avoid coupling multiple excitation lasers
to the same vibronic level, thus requiring separation of the main
cycling and repumping lasers to address different electronic levels
(e.g. main cycling on $\tilde{X}-\tilde{A}$ while repumping on $\tilde{X}-\tilde{B}$
or vise versa).

\begin{figure}
\begin{centering}
\includegraphics[width=8cm]{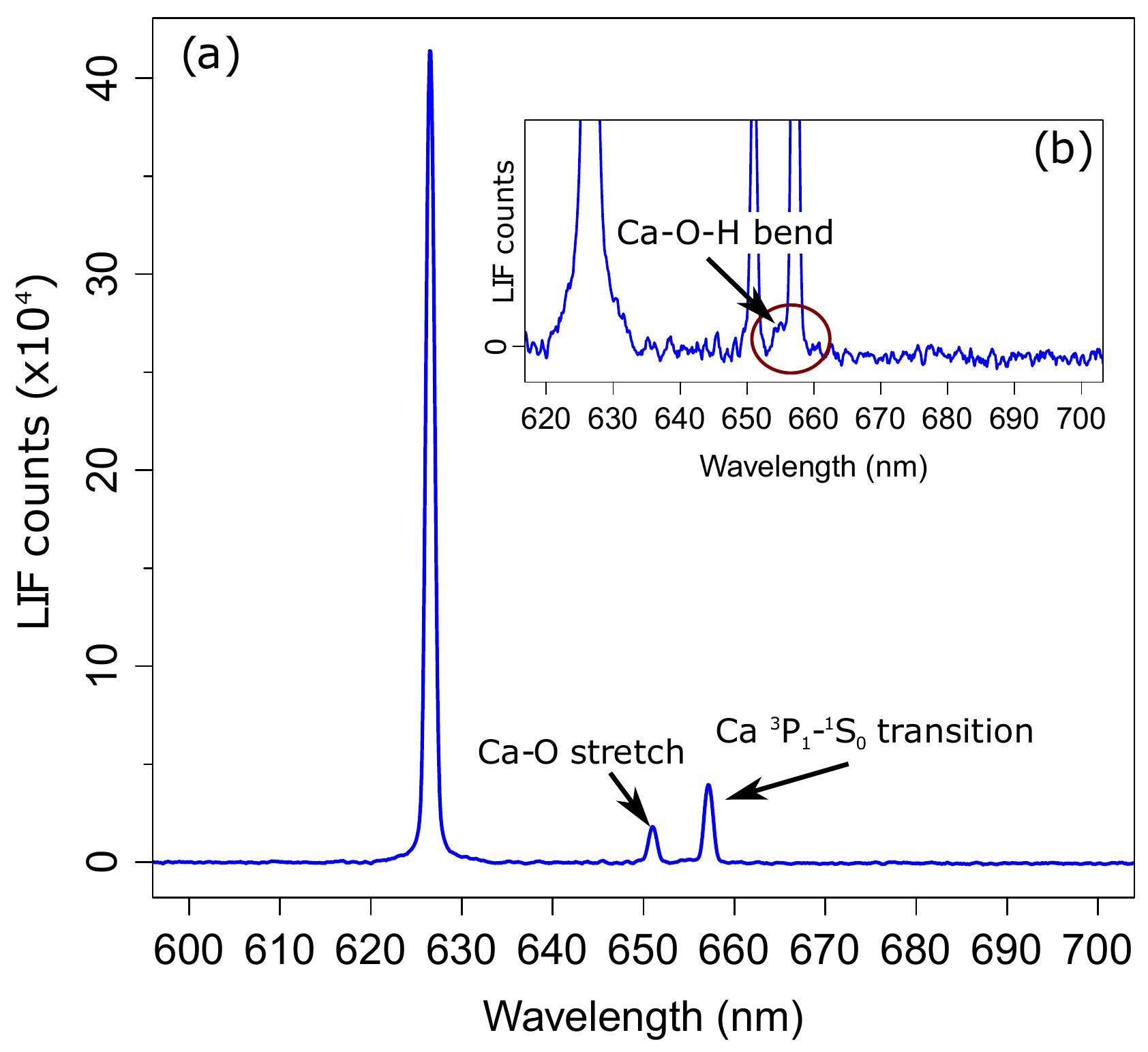} 
\par\end{centering}
\caption{\label{fig:CaOH_X2A}Measured CaOH vibrational branching in the $\tilde{X}-\tilde{A}$
electronic system. (a) Dispersed laser-induced fluorescence for CaOH
excited on the $P_{1}\left(N''=1\right)$ branch of the $\tilde{X}^{2}\Sigma^{+}\left(000\right)\rightarrow\tilde{A}^{2}\Pi_{1/2}\left(000\right)$
transition at 626 nm. $4.3\pm0.2\,\%$ of the molecules decay to the
first excited level of Ca-O stretching mode. The peak at 657 nm is
due to spontaneous emission on the intercombination $^{3}P_{1}\rightarrow^{1}S_{0}$
line of metastable calcium atoms created during the laser ablation
process. (b) Zoomed in region of the plot indicating that all other
decay channels are below 0.3\%. }
\end{figure}
\begin{figure}
\begin{centering}
\includegraphics[width=8cm]{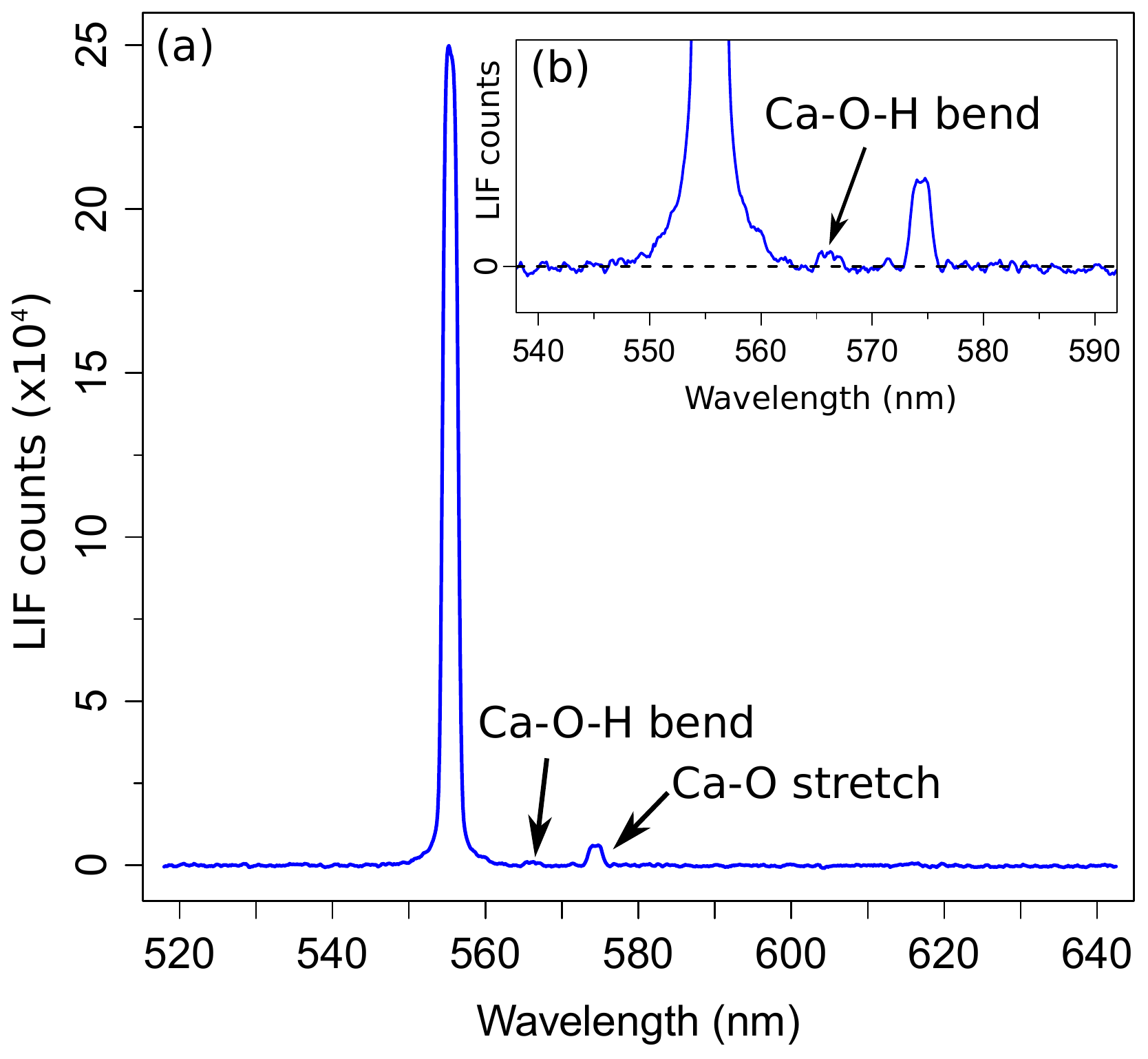} 
\par\end{centering}
\caption{\label{fig:(a)-Dispersed-LIF-CaOH_X2B}(a) Dispersed LIF data for
CaOH excited on the $\tilde{X}-\tilde{B}$ electronic transition at
555 nm. (b) Expanded 540-590 nm spectral region indicating a small
decay to Ca-O-H bending mode with 1 quanta. $\triangle v_{2}=1$ transitions
arising from intensity borrowing due to spin-orbit vibronic interaction
has been previously observed in the SrOH spectrum \cite{brazier1985laser}. }
\end{figure}
Figures \ref{fig:CaOH_X2A} and \ref{fig:(a)-Dispersed-LIF-CaOH_X2B}
display the results of the DLIF for CaOH excited using the $P_{1}\left(N''=1\right)$
transition on the $\tilde{X}-\tilde{A}^{2}\Pi_{1/2}$ (at 15964.38
cm\textsuperscript{-1}) and $\tilde{X}-\tilde{B}$ (at 18021.58 cm\textsuperscript{-1})
bands, correspondingly. Similarly to the previous measurements with
isoelectronic diatomic CaF \cite{wall2008lifetime} and triatomic
SrOH \cite{nguyen2018fluorescence}, the intensity of the off-diagonal
vibrational bands decreases rapidly, indicating the suitability of
using either electronic transition for optical cycling. High sensitivity
of our measurement allowed us to observe very weak decays to $\left(02^{0}0\right)$
and $\left(01^{1}0\right)$ excited bending vibrational states as
shown in Fig. \ref{fig:CaOH_X2A}(b) and \ref{fig:(a)-Dispersed-LIF-CaOH_X2B}(b).
Symmetry forbidden $\triangle l\neq0$ vibronic decays in alkaline
earth monohydroxides occur in second order due to $H_{{\rm RT}}\times H_{{\rm SO}}$
Hamiltonian term. The dipolar term of the Renner-Teller (RT) perturbation
operator connects basis functions with $v_{2}=\pm1$ and $\triangle l=-\triangle\Lambda=\pm1$
\cite{bolman1973renner,brown1977effective} and together with the
off-diagonal parts of the spin-orbit (SO) operator leads to coupling
between $\tilde{B}\left(000\right)\sim\tilde{B}\left(01^{1}0\right)$
states \cite{presunka1994laser}. Decay to the $\left(02^{0}0\right)$
vibrational state is symmetry allowed and observed in our measurements
(Fig. \ref{fig:CaOH_X2A}(b)), but the more precise determination
of this band's intensity is limited by the contamination signal arising
due to the 657 nm emission from the metastable atomic calcium created
during the laser ablation process.

\subsection{CaOCH\protect\protect\textsubscript{3} measurements}

\begin{figure}
\centering{}\includegraphics[width=8cm]{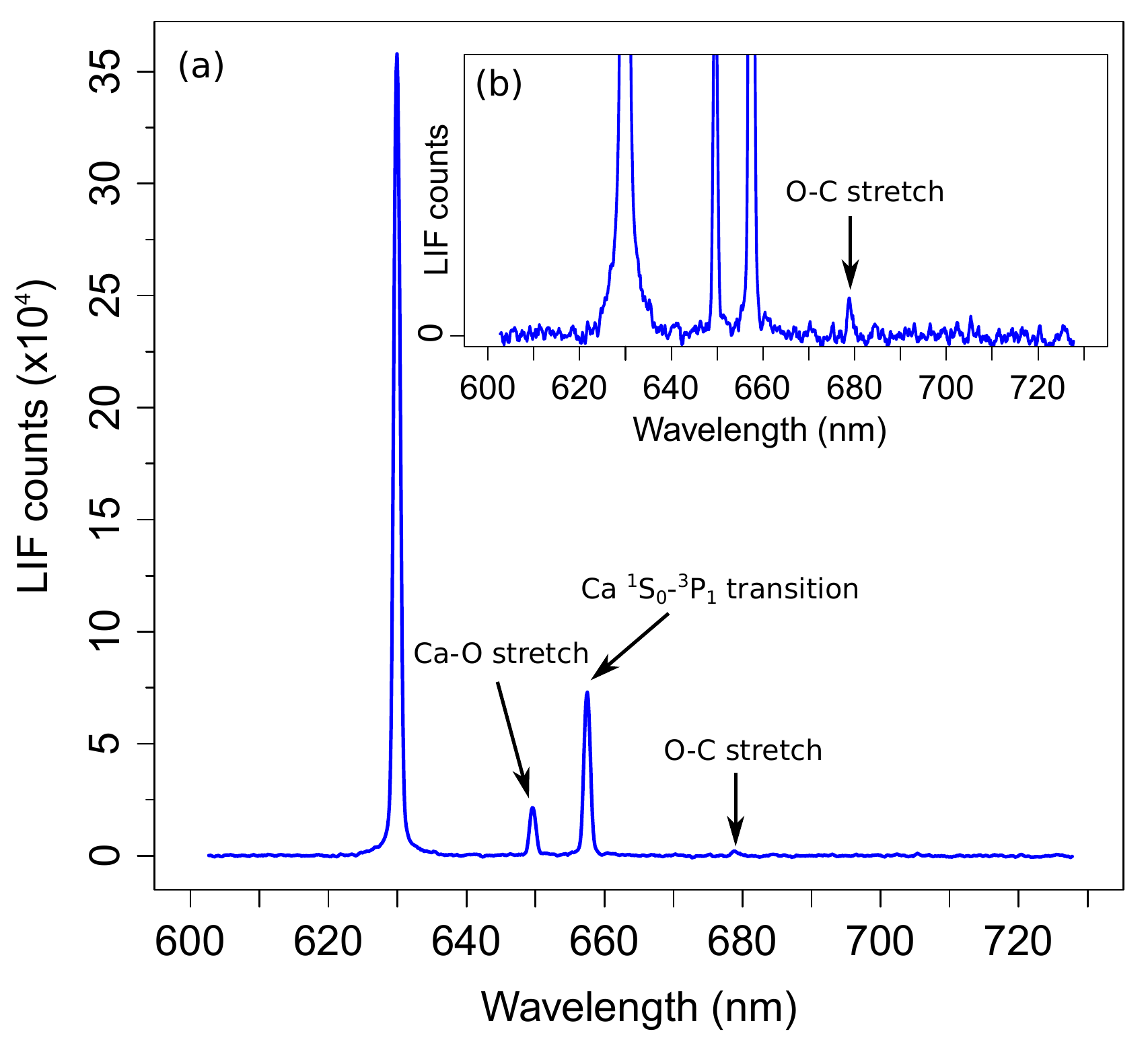}\caption{\label{fig:CaOCH3_X2A}Measurement of the CaOCH\protect\protect\textsubscript{3}
vibrational branching in the $\tilde{X}-\tilde{A}$ electronic system
following excitation on the $P_{1}\left(N''=1\right)$,  $K''=1\rightarrow K'=0$
transition at 15886.18 cm\protect\textsuperscript{-1}. Like in the
CaOH data, there is a residual signal at 657 nm coming from the spontaneous
emission on the intercombination $^{3}P_{1}\rightarrow^{1}S_{0}$
line of metastable calcium atoms. Plot (a) shows that the dominant
off-diagonal decay is to the excited Ca-O stretching vibration with
one quanta. Plot (b) demonstrates a small peak at 679 nm due to the
decay into the excited totally-symmetric vibrational mode composed
of O-C and C-H stretching vibrational motions. }
\end{figure}

The presence of $K\neq0$ states for non-linear molecules requires
additional considerations for achieving effective photon cycling beyond
those examined previously for diatomic and linear polyatomic molecules.
Reference \cite{Kozyryev2016MOR} outlined how to use $K''=1\rightarrow K'=0$
transition for the perpendicular $\tilde{X}-\tilde{A}$ transition
and $K''=0\rightarrow K'=0$ transition for the parallel $\tilde{X}-\tilde{B}$
to achieve rotationally closed excitations. Figures \ref{fig:CaOH_X2A}
and \ref{fig:CaOCH3_X2A} provide a comparison of the dispersed laser
induced fluorescence for CaOH and CaOCH\textsubscript{3}. Both of
the molecules were excited on the $\tilde{X}-\tilde{A}$ electronic
transitions around 630 nm and in a rotationally resolved manner to
$J'=1/2$ state. As can be clearly seen from the data presented, both
CaOH and CaOCH\textsubscript{3} have a very small number of vibrational
decay channels despite significantly increased structural complexity
in going from a triatomic to a hexatomic molecule. While CaOCH\textsubscript{3}
has eight distinct vibrational modes (4 of $a_{1}$ symmetry and 4
degenerate pairs of $e$ symmetry), only two are optically active
at the three part per thousand level as seen in the figure. Our measurements,
performed on a cold molecular beam with a narrow-band dye laser exciting
only the rotational level suitable for laser cooling, confirm that
even non-linear CaOR molecules behave effectively like triatomics,
showing great promise for laser cooling and trapping.

\begin{figure}
\begin{centering}
\includegraphics[width=8cm]{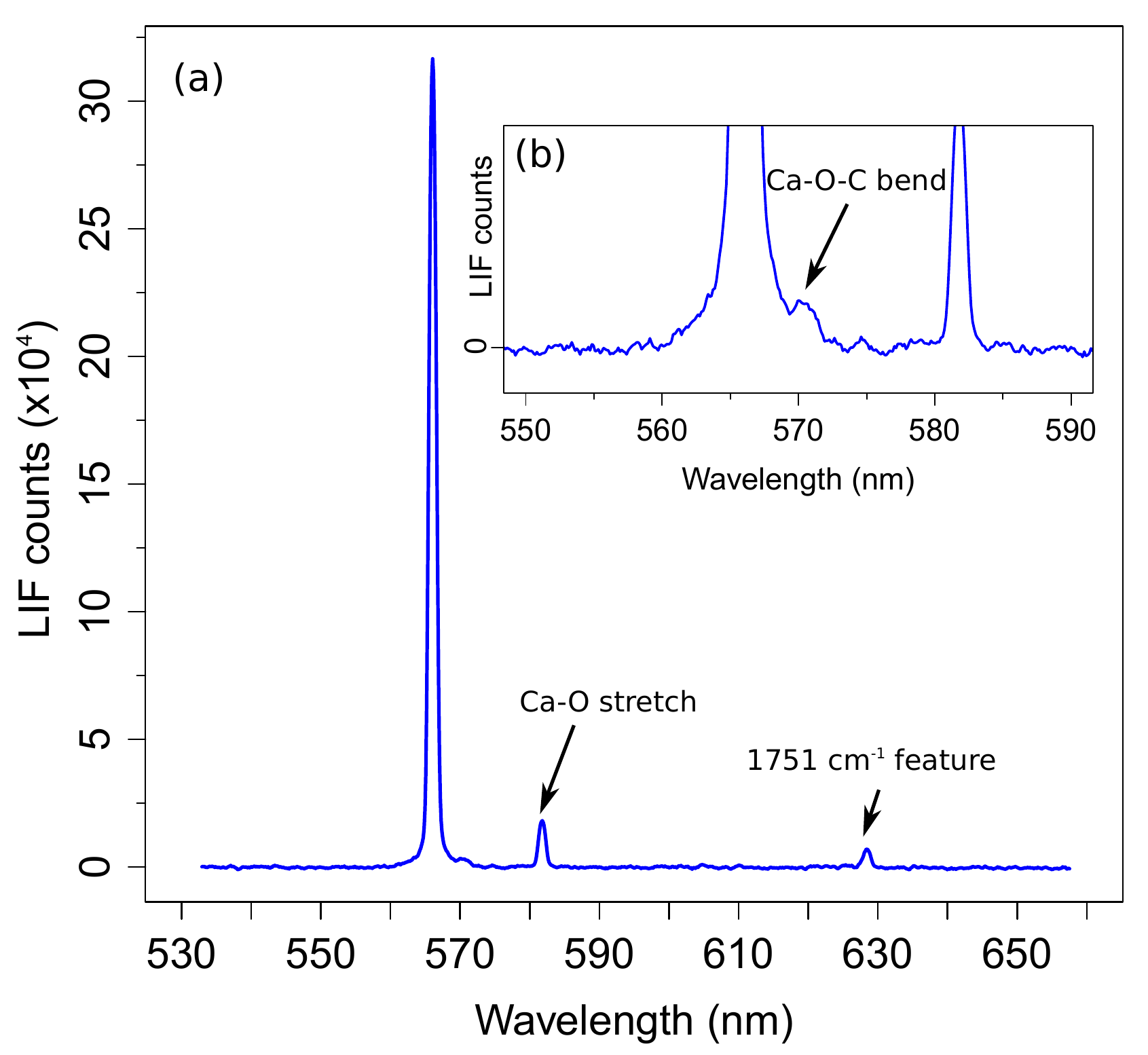} 
\par\end{centering}
\caption{\label{fig:Dispersed-LIF-X2B-methoxy}Dispersed laser-induced fluorescence
data for CaOCH\protect\protect\textsubscript{3} excited on the $\tilde{X}-\tilde{B}$
electronic transition at 566 nm using $P_{1}\left(N''=1\right)$,
$K''=0\rightarrow K'=0$ branch at 17682.20 cm\protect\textsuperscript{-1}.
(a) Besides the dominant decay to the Ca-O stretching mode, an additional
unexpected feature around 1750 cm\protect\textsuperscript{-1} is
observed. (b) Lineshape assymetry on the main peak indicated a $\triangle l\protect\neq0$
decay to the Ca-O-C bending mode with one unit of vibrational angular
momentum $l$. }
\end{figure}
\begin{figure}
\centering{}\includegraphics[width=9cm]{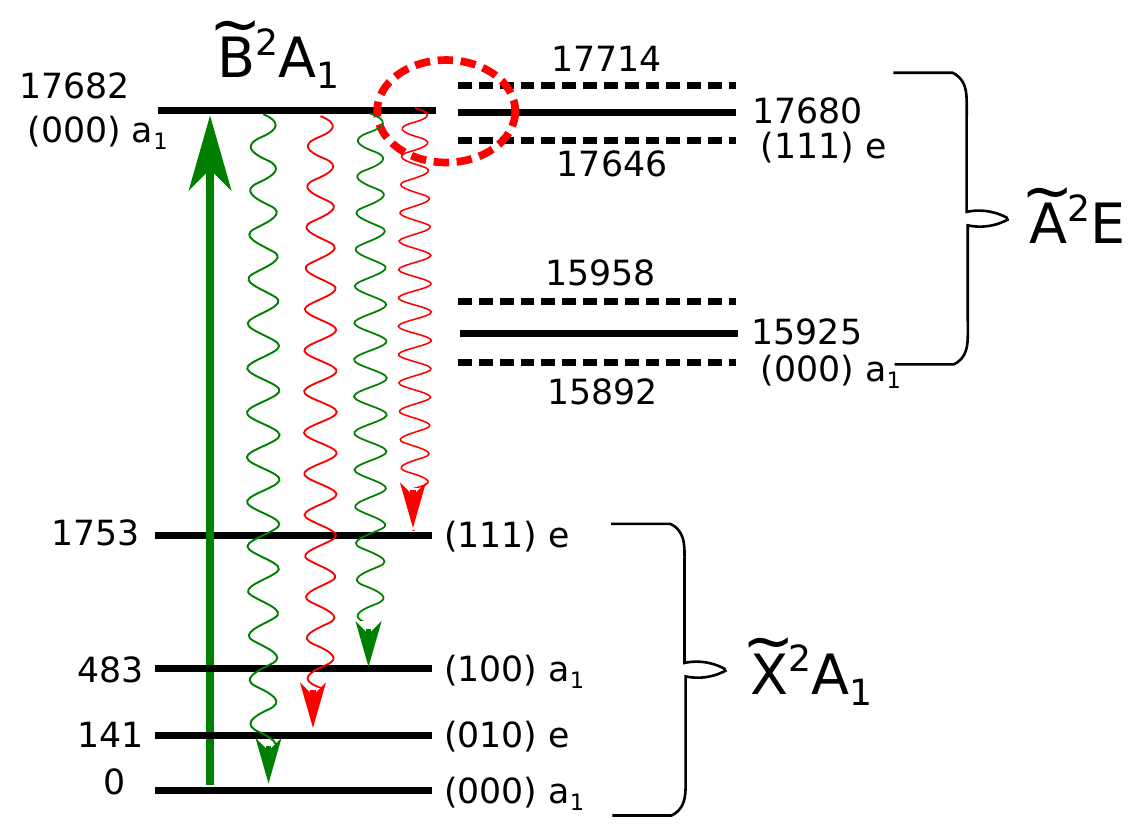}\caption{\label{fig:Proposed-mechanism-of-PJT}A proposed mechanism of vibronic
coupling in CaOCH\protect\protect\protect\textsubscript{3} leading
to symmetry-forbidden vibrational emissions. By analogy with CaOH,
triatomic notation is used for annotating the vibrational modes: ($v_{1}$00)
= Ca-O stretch, (0$v_{2}$0) = Ca-O-C bend, and (00$v_{3}$) = O-C
stretch. Red arrows indicate symmetry-forbidden decays allowed by
the vibronic mixing between $\tilde{B}\left(000\right)$ and $\tilde{A}\left(111\right)$.
Dashed energy levels indicate the effects of the spin-orbit splitting
in the $\tilde{A}$ state.}
\end{figure}

We have also used dispersed LIF to explore the possibility of using
the $\tilde{X}-\tilde{B}$ electronic transition in CaOCH\textsubscript{3}
around 566 nm for laser cooling. Figure \ref{fig:Dispersed-LIF-X2B-methoxy}
shows the DLIF results indicating two additional features not present
in the $\tilde{A}\rightarrow\tilde{X}$ emission spectrum: i) emission
to 1 quanta of the Ca-O-C bending mode and ii) a new feature around
1750 cm\textsuperscript{-1} away from the excitation band. Since
$\tilde{B}$ is the second excited electronic level, there is possibility
of mixing with the excited vibrational levels of the $\tilde{A}$
state. The proposed intensity borrowing mechanism is shown in Fig.
\ref{fig:Proposed-mechanism-of-PJT}, indicating the resulting decays
nominally forbidden by the symmetry arguments. As clearly seen in
the data presented in Fig. \ref{fig:Dispersed-LIF-X2B-methoxy}, a
relatively strong emission to the band offset by 1751 cm\textsuperscript{-1}
from the diagonal vibronic transition is observed in the $\tilde{B}\rightarrow\tilde{X}$
decay of CaOCH\textsubscript{3}, making laser cooling on this electronic
transition more challenging compared to $\tilde{X}\rightarrow\tilde{A}^{2}E_{1/2}$. 

Any nonlinear molecule (e.g. CaOCH\textsubscript{3}) in an orbitally
degenerate electronic state (e.g. $^{2}E$ state) will always distort
in a way such as to lower the symmetry and remove degeneracy. This
so-called Jahn-Teller effect (JTE) nullifies the $\triangle v_{i}=\pm2,\pm4,\ldots$
selection rule for nonsymmetric vibrations in electronic transitions
\cite{bernath2005spectra}. Jahn-Teller effect in the $\tilde{A}$
electronic state as well as near-degeneracy between the excited combination
band in the $\tilde{A}$ state and the ground vibrational level of
the $\tilde{B}$ electronic state lead to Born-Oppenheimer approximation
breakdown. Since the eigenstates of the new Hamiltonian are a mixture
of the unperturbed eigenstates of the $\tilde{B}\left(000\right)$
and $\tilde{A}\left(111\right)$ levels, forbidden decays are observed
in the emission spectrum. The appearance of the pseudo-JTE in the
presence of accidentally near-degenerate states (as depicted in Fig.
\ref{fig:Proposed-mechanism-of-PJT}) is carefully described in Ref.
\cite{fischer1984vibronic}. The perturbing operator of the polynomial
form in the normal coordinate $Q_{i}$ about the molecular configuration
$Q_{0}$ \cite{fischer1984vibronic} 
\begin{equation}
H'=\left(\frac{\partial H}{\partial Q_{i}}\right)_{Q_{0}}Q_{i}+\frac{1}{2}\left(\frac{\partial^{2}H}{\partial Q_{i}^{2}}\right)_{Q_{0}}Q_{i}^{2}+\ldots
\end{equation}
is responsible for mixing of the zeroth-order Born-Oppenheimer vibrational
and electronic wavefunctions. Careful theoretical understanding of
vibronic coupling in the excited $\tilde{B}$ state is currently in
progress and beyond the scope of the present paper. 

\section{comparison with theoretical predictions}

\begin{table}
\begin{centering}
\begin{tabular}{cccccc}
\hline 
State  & Mode (symmetry)  & TI  & TD  & Measured \cite{jacox1998PolyA}  & Ref. \cite{isaev2015polyatomic}\tabularnewline
\hline 
$\tilde{X}$  & bend ($\Pi$)  & 356.72  & 344.89/345.29  & 352.93  & 411/408\tabularnewline
$\tilde{X}$  & Ca-O stretch ($\Sigma$)  & 625.22  & 603.84  & 609.02  & 622\tabularnewline
$\tilde{X}$  & O-H stretch ($\Sigma$)  & 3816.28  & 3819.96  & 3778  & 4276\tabularnewline
$\tilde{A}$  & bend ($\Pi$)  & 368.41  & 370.37/371.13  & 361.36  & 402/386\tabularnewline
$\tilde{A}$  & Ca-O stretch ($\Sigma$)  & 625.18  & 603.84  & 630.68  & 646\tabularnewline
$\tilde{A}$  & O-H stretch ($\Sigma$)  & 3816.06  & 3819.96  & -  & 4279\tabularnewline
\hline 
\end{tabular}
\par\end{centering}
\caption{\label{tab:normal-modes-CaOH}Calculated and measured normal vibrational
modes (in units of cm\protect\protect\textsuperscript{-1}) for CaOH.
TI and TD stand for time-independent and time-dependent density functional
theory calculations, respectively. }
\end{table}
\begin{table}
\begin{centering}
\begin{tabular}{ccccc}
\hline 
State  & Mode (symmetry)  & TI  & TD  & Measured \cite{jacox2003PolyB}\tabularnewline
\hline 
$\tilde{X}$  & Ca-O-C bend ($e$)  & 138.59/140.51  & 133.11/142.09  & $144\pm5$\tabularnewline
$\tilde{X}$  & Ca-O stretch ($a_{1}$)  & 482.90  & 472.09  & $488\pm5$\tabularnewline
$\tilde{X}$  & Sym. CH\textsubscript{3} bend \& C-O stretch ($a_{1}$)  & 1132.20  & 945.19  & $1156\pm5$\tabularnewline
$\tilde{X}$  & asymmetric combination ($e$)  & 1134.30/1134.61  & 1109.93/1120.49  & -\tabularnewline
$\tilde{X}$  & Sym. CH\textsubscript{3} bend \& C-O stretch ($a_{1}$)  & 1425.76  & 1385.36  & -\tabularnewline
$\tilde{X}$  & asymmetric combination ($e$)  & 1444.06/1444.42  & 1425.37/1434.73  & -\tabularnewline
$\tilde{X}$  & Sym. CH\textsubscript{3} stretch ($a_{1}$)  & 2879.18  & 2886.36  & -\tabularnewline
$\tilde{X}$  & asymmetric stretch ($e$)  & 2922.11/2922.61  & 2933.60/2933.75  & -\tabularnewline
\hline 
\end{tabular}
\par\end{centering}
\caption{\label{tab:normal-modes-CaOCH3-X}Calculated and measured normal vibrational
modes (in units of cm\protect\protect\protect\textsuperscript{-1})
for the $\tilde{X}$ electronic state of CaOCH\protect\protect\protect\textsubscript{3}.}
\end{table}
\begin{table}
\begin{centering}
\begin{tabular}{ccccc}
\hline 
State  & Mode (symmetry)  & TI  & TD  & Measured \cite{jacox2003PolyB}\tabularnewline
\hline 
$\tilde{A}$  & Ca-O-C bend ($e$)  & 141.43/141.54  & 129.76/141.42  & $145\pm5$\tabularnewline
$\tilde{A}$  & Ca-O stretch ($a_{1}$)  & 482.66  & 472.13  & 501.48\tabularnewline
$\tilde{A}$  & Sym. CH\textsubscript{3} bend \& C-O stretch ($a_{1}$)  & 1132.24  & 1119.84  & $1140\pm5$\tabularnewline
$\tilde{A}$  & asymmetric combination ($e$)  & 1134.54/1134.62  & 1206.80/1269.47  & -\tabularnewline
$\tilde{A}$  & Sym. CH\textsubscript{3} bend \& C-O stretch ($a_{1}$)  & 1425.48  & 1425.10  & -\tabularnewline
$\tilde{A}$  & asymmetric combination ($e$)  & 1444.08/1444.13  & 1672.82/1741.32  & -\tabularnewline
$\tilde{A}$  & Sym. CH\textsubscript{3} stretch ($a_{1}$)  & 2879.54  & 2886.29  & -\tabularnewline
$\tilde{A}$  & asymmetric stretch ($e$)  & 2923.32/2923.36  & 2931.02/2934.70  & -\tabularnewline
\hline 
\end{tabular}
\par\end{centering}
\caption{\label{tab:Calculated-and-measured-normal_modes_CAOR_A}Calculated
and measured normal vibrational modes (in units of cm\protect\protect\protect\textsuperscript{-1})
for the $\tilde{A}$ electronic state of CaOCH\protect\protect\protect\textsubscript{3}.}
\end{table}
Because of the highly diagonal Franck-Condon factor matrices for both
CaOH and CaOCH\textsubscript{3} molecules, only a few excited vibrational
modes were present in our DLIF data as seen in Fig. \ref{fig:CaOH_X2A}-\ref{fig:Dispersed-LIF-X2B-methoxy}.
Other than the unexpected peak seen in the 1751 nm overtone band for
the $\tilde{B}\rightarrow\tilde{X}$ emission, the frequencies of
other vibrational modes have been previously measured and tabulated
which allowed for unambiguous vibrational character assignment. Theoretical
analysis of the multidimensional Franck-Condon factors for all vibrational
modes -- including those not observed in the experiment -- depends
on the frequencies of the corresponding vibrational motions. We therefore
initialized our calculations with ab initio molecular geometries and
vibrational frequencies using the ORCA quantum chemistry program,
which is discussed in Ref. \cite{neese2012orca}. Tables \ref{tab:normal-modes-CaOH},
\ref{tab:normal-modes-CaOCH3-X} and \ref{tab:Calculated-and-measured-normal_modes_CAOR_A}
provide a summary of the calculated normal vibrational modes for CaOH
and CaOCH\textsubscript{3}. We observe excellent agreement between
our ORCA calculations with previous ab initio calculations for CaOH
by Isaev and Berger \cite{isaev2015polyatomic} as well as previously
experimentally determined values for the vibrational normal modes.
In the case of MOCH$_{3}$ molecules, there are 12 normal modes, 4
of $a_{1}$ symmetry and 4 degenerate pairs of $e_{1}$ symmetry.
Thus, while there are 12 internal coordinates, we reduced the problem
to 8 symmetry coordinates listed in Table \ref{tab:Symmetry-coordinates-for-MOR}. 

\begin{table}
\begin{centering}
\begin{tabular}{ccc}
\hline 
 & Coordinate & Symmetry\tabularnewline
\hline 
1 & CH\textsubscript{3} symmetric stretch & $a_{1}$\tabularnewline
2 & CH\textsubscript{3} symmetric bend & $a_{1}$\tabularnewline
3 & C-O stretch & $a_{1}$\tabularnewline
4 & M-O stretch & $a_{1}$\tabularnewline
5 & CH\textsubscript{3} asymmetric stretch & $e$\tabularnewline
6 & CH\textsubscript{3} asymmetric bend & $e$\tabularnewline
7 & O-CH\textsubscript{3} wag & $e$\tabularnewline
8 & M-O-C bend & $e$\tabularnewline
\hline 
\end{tabular}
\par\end{centering}
\caption{\label{tab:Symmetry-coordinates-for-MOR}Symmetry coordinates for
MOCH$_{3}$ molecules, listed in an order identical to the basis used
for GF calculations. Lower case symmetries ($a_{1}$, $e$) correspond
to the irreducible representations of the C$_{3v}$ point group. Coordinates
are labeled for the normal modes to which they roughly correspond.}

\end{table}
The details of the numerical computation of CaOH and CaOCH\textsubscript{3}
Franck-Condon factors are described in the Appendix. The vibrational
branching ratios were computed by weighting each FCF by the cube of
the transition frequency from the vibrationless excited electronic
state to the vibrational state in the ground electronic state and
normalizing. Results of these calculations are shown in Tables \ref{tab:CaOH_measured_branching}
and \ref{tab:CaOCH3_measured_branching}. We obtain excellent agreement
between the measured and calculated vibrational branching ratios for
CaOH and a reasonable agreement for a more complex CaOCH\textsubscript{3}
molecule. Incorporating possible couplings between different vibrational
modes as well as inclusion of the anharmonic terms in the vibrational
potential could resolve the slight discrepancy between the measured
and predicted vibrational branching ratios for CaOCH\textsubscript{3}.
Notice that our theoretical predictions for FCFs are much closer to
the measured values we have observed compared to the ab initio values
predicted by Isaev and Berger \cite{isaev2015polyatomic}. By using
available experimental vibrational frequencies to determine the force
constants as well as incorporating experimentally measured bond lengths
and structure into our analysis, we were able to circumvent some of
the challenges associated with predictions of FCFs with sufficient
accuracy for designing a molecular laser cooling scheme \cite{isaev2010laser}.
As seen in Table \ref{tab:CaOH_measured_branching}, purely ab initio
calculations for even diatomic FCFs can potentially underestimate
the feasibility of direct laser cooling, while calculations relying
on molecular parameters extracted from previous spectroscopic data
can provide a more accurate guide to developing a successful laser
cooling experiment for a new molecule with desired properties (e.g.
new physics sensitivity \cite{norrgard2018nuclear}). 

\begin{table}
\begin{centering}
\begin{tabular}{cccccc}
\hline 
Band  & $\lambda_{v',v''}$, nm  & Calc. FCFs  & Ref. \cite{isaev2015polyatomic} FCFs  & Calc. VBRs  & Obs. VBRs\tabularnewline
\hline 
$0_{0}^{0}\tilde{A}^{2}\Pi_{1/2}-\tilde{X}^{2}\Sigma^{+}$  & 626.5  & 0.9521  & 0.9213  & 0.9570  & $0.957\pm0.002$\tabularnewline
$1_{1}^{0}\tilde{A}^{2}\Pi_{1/2}-\tilde{X}^{2}\Sigma^{+}$  & 651.0  & 0.0459  & 0.0763  & 0.0410  & $0.043\pm0.002$\tabularnewline
$2_{2}^{0}\tilde{A}^{2}\Pi_{1/2}-\tilde{X}^{2}\Sigma^{+}$  & 654.3  & $3\times10^{-4}$  & -  & $3\times10^{-4}$  & $3_{-2}^{+1}\times10^{-3}$\tabularnewline
$0_{0}^{0}\tilde{A}^{2}\Pi_{3/2}-\tilde{X}^{2}\Sigma^{+}$  & 623.9  & 0.9521  & 0.9213  & 0.959 & $0.959\pm0.005$\tabularnewline
$1_{1}^{0}\tilde{A}^{2}\Pi_{3/2}-\tilde{X}^{2}\Sigma^{+}$  & 648.3  & 0.0459  & 0.0763  & 0.041 & $0.041\pm0.005$\tabularnewline
$0_{0}^{0}\tilde{B}^{2}\Sigma^{+}-\tilde{X}^{2}\Sigma^{+}$  & 555.2  & 0.9711  & -  & 0.9742  & $0.975\pm0.001$\tabularnewline
$1_{1}^{0}\tilde{B}^{2}\Sigma^{+}-\tilde{X}^{2}\Sigma^{+}$  & 574.4  & 0.0270  & -  & 0.0244  & $0.022\pm0.001$\tabularnewline
$2_{1}^{0}\tilde{B}^{2}\Sigma^{+}-\tilde{X}^{2}\Sigma^{+}$  & 566.2  & 0  & -  & 0  & $0.003\pm0.001$\tabularnewline
\hline 
\end{tabular}
\par\end{centering}
\caption{\label{tab:CaOH_measured_branching}Summary of the measured and calculated
branching ratios for CaOH molecule.}
\end{table}
\begin{table}
\begin{centering}
\begin{tabular}{ccccc}
\hline 
Band  & $\lambda_{v',v''}$, nm  & Calc. FCFs  & Calc. VBRs  & Obs. VBRs\tabularnewline
\hline 
$0_{0}^{0}\tilde{A}^{2}E_{1/2}-\tilde{X}^{2}A_{1}$  & 629.9  & 0.944  & 0.950  & $0.931\pm0.003$\tabularnewline
$1_{1}^{0}\tilde{A}^{2}E_{1/2}-\tilde{X}^{2}A_{1}$  & 649.5  & 0.050  & 0.045  & $0.063\pm0.003$\tabularnewline
$3_{1}^{0}\tilde{A}^{2}E_{1/2}-\tilde{X}^{2}A_{1}$  & 678.8  & 0.003  & 0.002  & $0.006\pm0.003$\tabularnewline
$0_{0}^{0}\tilde{A}^{2}E_{3/2}-\tilde{X}^{2}A_{1}$  & 627.1  & 0.944  & 0.954 & $0.945\pm0.004$\tabularnewline
$1_{1}^{0}\tilde{A}^{2}E_{3/2}-\tilde{X}^{2}A_{1}$  & 646.6  & 0.050  & 0.046 & $0.055\pm0.004$\tabularnewline
$0_{0}^{0}\tilde{B}^{2}A_{1}-\tilde{X}^{2}A_{1}$  & 565.9  & 0.941  & 0.947  & $0.910\pm0.002$\tabularnewline
$1_{1}^{0}\tilde{B}^{2}A_{1}-\tilde{X}^{2}A_{1}$  & 581.6  & 0.049  & 0.045  & $0.057\pm0.002$\tabularnewline
$2_{1}^{0}\tilde{B}^{2}A_{1}-\tilde{X}^{2}A_{1}$  & 570.3  & 0  & 0  & $0.006\pm0.002$\tabularnewline
$1_{1}^{0}2_{1}^{0}3_{1}^{0}\tilde{B}^{2}A_{1}-\tilde{X}^{2}A_{1}$  & 628.2  & 0  & 0  & $0.026\pm0.002$\tabularnewline
\hline 
\end{tabular}
\par\end{centering}
\caption{\label{tab:CaOCH3_measured_branching}Summary of the measured and
calculated branching ratios for CaOCH\protect\protect\textsubscript{3}
molecule. }
\end{table}

\section{laser cooling and trapping prospects}

\begin{figure}
\begin{centering}
\includegraphics[width=10cm]{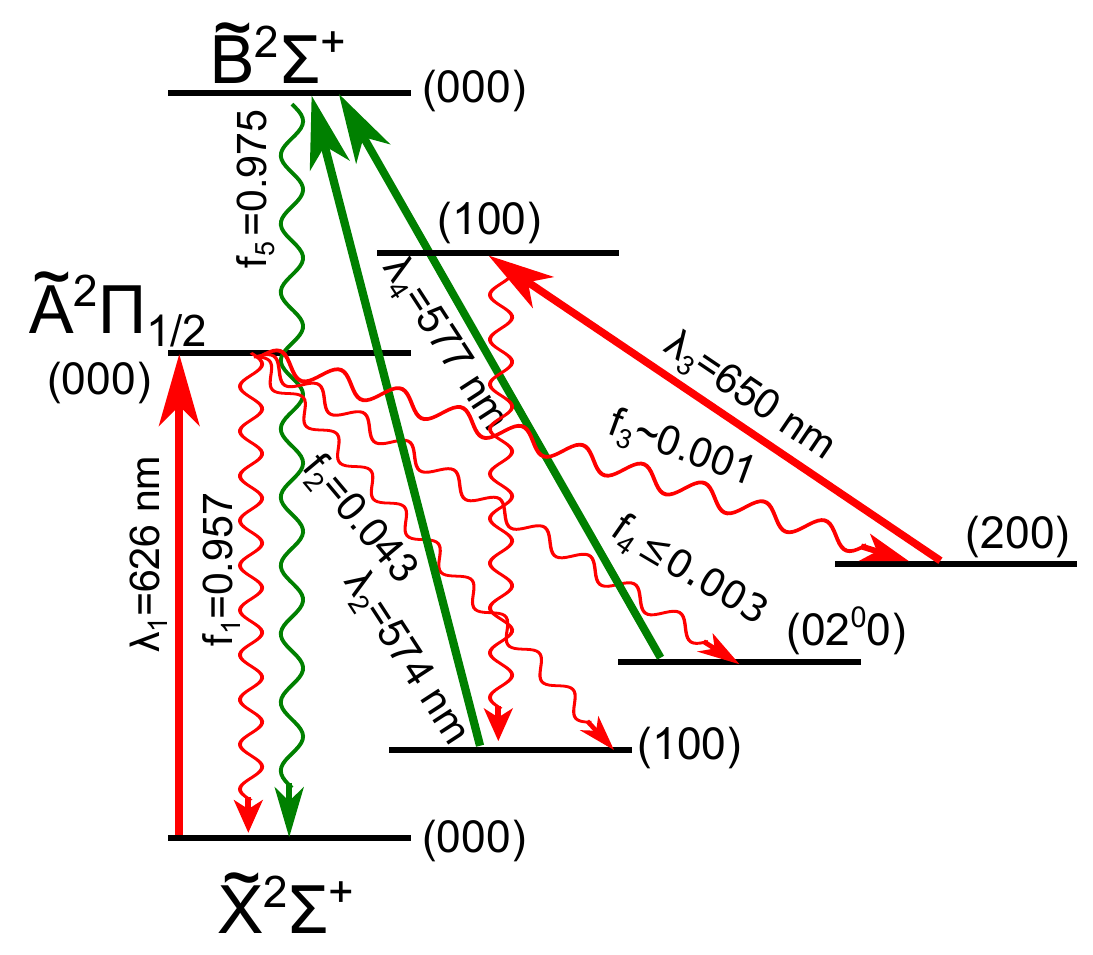} 
\par\end{centering}
\caption{\label{fig:Proposed-laser-cooling-CaOH}Proposed laser cooling and
trapping scheme for CaOH based on the experimental results. }
\end{figure}

\begin{figure}
\begin{centering}
\includegraphics[width=12cm]{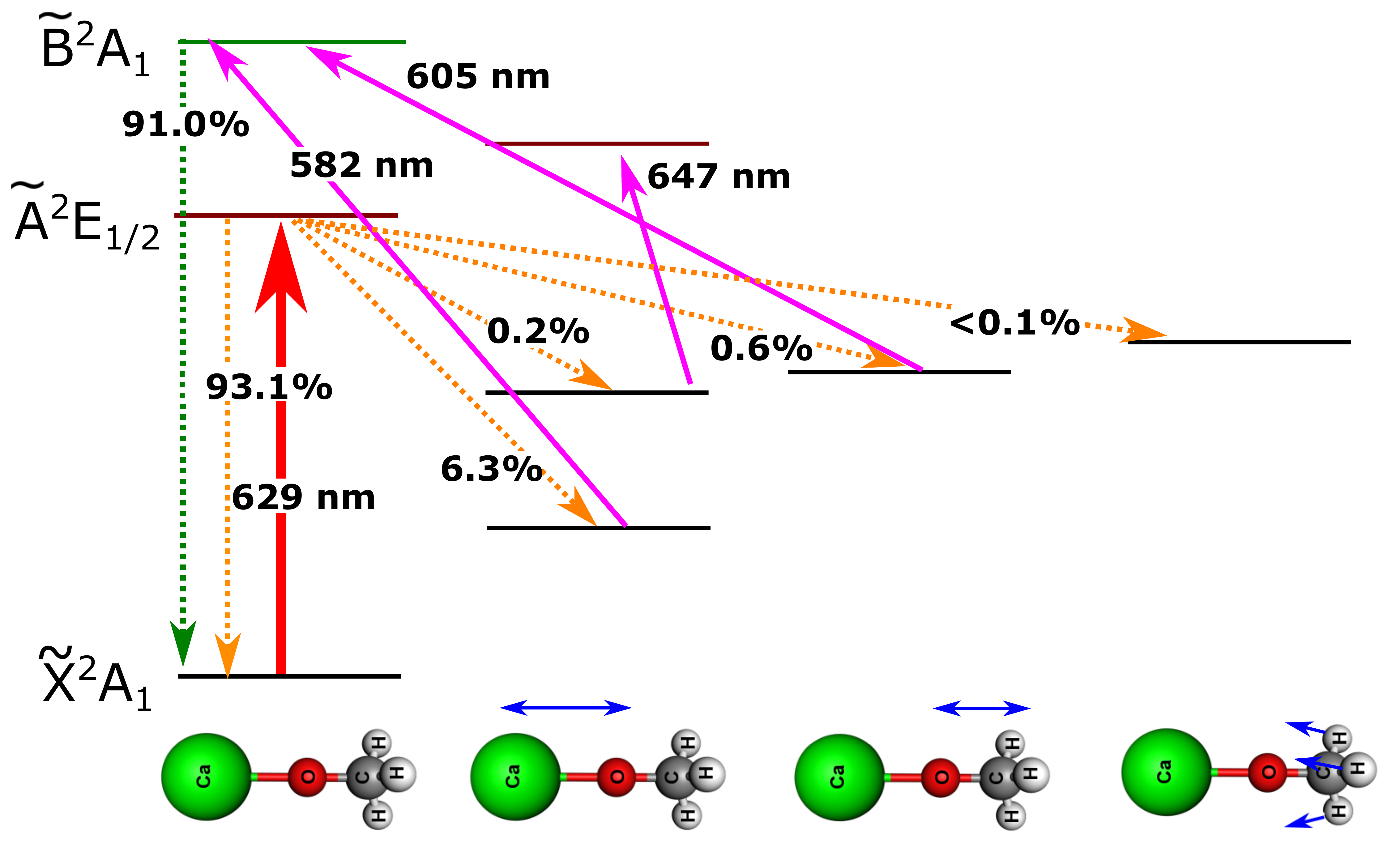} 
\par\end{centering}
\caption{\label{fig:Measured-vibrational-branching}Measured vibrational branching
ratios for CaOCH\protect\protect\textsubscript{3} along with proposed
laser cooling and repumping scheme. The dominant character of the
appropriate vibrational mode is indicated.}
\end{figure}

Our experimental measurements and theoretical calculations confirm
the possibility of achieving direct laser cooling for CaOH and CaOCH\textsubscript{3}
with a few repumping lasers. Furthermore, our measurements provide
the quantitative guidance to choosing the most experimentally efficient
route to identifying the main cooling and all the repumping transition
bands. Using the measured values for the branching ratios, we have
determined the optimal laser cooling schemes for CaOH and CaOCH\textsubscript{3}
depicted in Fig. \ref{fig:Proposed-laser-cooling-CaOH} and \ref{fig:Measured-vibrational-branching},
respectively. Because of the relatively light mass, CaOH can be slowed
down to a complete stop from 100 m/s cryogenic buffer-gas beam using
only 9,000 photons which can be obtained with three repumping lasers
indicated. Thus, the technical complexity of laser slowing CaOH is
comparable to experiments with diatomic molecules SrF and CaF. 

While additional structural complexity for CaOCH\textsubscript{3}
results in less diagonal Franck-Condon factors, with the repumping
scheme in Fig. \ref{fig:Measured-vibrational-branching} and assuming
$\sim10^{9}$ molecules in a single rovibrational quantum state at
100 m/s, one can obtain between $10^{4}$ and $10^{7}$ slowed molecules
in the trapping region, assuming 1 part per 1,000 or 1 part per 3,000
loss probability to unaddressed vibrational levels. 

\section{beyond symmetric top molecules}

Because of the strongly ionic nature of the Ca-O bond in CaOR molecules,
the electronic transitions involve non-bonding Ca\textsuperscript{+}
orbitals perturbed by an RO\textsuperscript{- } ligand even for asymmetric
rotor molecules (ARMs). Since the Ca-O-C bond angle remains $180^{\circ}$,
the local symmetry near the calcium metal is linear and therefore
many characteristics of electronic transitions carry over from linear
CaOH to more complex CaOR symmetric top and asymmetric rotor radicals
\cite{brazier1986laser,brazier1985organometal}. For example, there
is ``spin-orbit'' splitting of about the same size, and the electronic
orbital angular momentum remains essentially unquenched. However,
the symmetry reduction in going from $C_{\infty v}$ (CaOH) and $C_{3v}$
(CaOCH\textsubscript{3}) to molecules with lower symmetry properties
like $C_{S}$ and no symmetry has potential to impact the vibrational
branching ratios in a significant manner. For linear ($C_{\infty v}$)
and symmetric top ($C_{3v}$) molecules, Ca-O-C bending vibrations
have $\pi$ or $e$ symmetry, correspondingly, and therefore only
$\triangle v=\pm2,\,\pm4,\ldots$ branches appear in the electronic
decays under the Born-Oppenheimer approximation \footnote{Weak forbidden vibronic transitions have been observed for CaOH and
SrOH enabled by spin-orbit interaction and Renner-Teller coupling.}. Such a bending mode becomes non-degenerate for molecules with lower
symmetries than $C_{3v}$, therefore leading to allowed decays of
comparable intensity to both one quanta in the Ca-O stretching and
Ca-O-C bending modes as has been previously observed for calcium monoalkoxides
\cite{brazier1986laser}. While this presents a technical challenge,
requiring an additional laser for repumping lost molecules from such
a decay channel, it does not pose a fundamental threat to achieving
photon cycling in asymmetric rotor molecules like CaOC\textsubscript{2}H\textsubscript{5}
or CaOCH\textsubscript{2}CH\textsubscript{3}.

Depending on the orientation of the transition dipole moment relative
to the principal axes of the molecule, vibrational and electronic
spectra of the ARMs are characterized as $a$-type, $b$-type or $c$-type
transitions with the selection rules \cite{bernath2005spectra}: 
\begin{itemize}
\item $a$-type bands: $\triangle K_{a}=0$, $\triangle K_{c}=\pm1$ and
$\triangle J=0,\,\pm1$, except for $K_{a}^{'}=0\leftarrow K_{a}^{''}=0$
for which $\triangle J=\pm1$ 
\item $b$-type bands: $\triangle K_{a}=\pm1$, $\triangle K_{c}=\pm1$
and $\triangle J=0,\,\pm1$ 
\item $c$-type bands: $\triangle K_{c}=0$, $\triangle K_{a}=\pm1$ and
$\triangle J=0,\,\pm1$, except for $K_{c}^{'}=0\leftarrow K_{c}^{''}=0$
for which $\triangle J=\pm1$. 
\end{itemize}
Therefore, as can be seen from the selection rules above, for near-oblate
or near-prolate ARMs, the overall structure of the emission bands
resembles that of parallel or perpendicular transitions for STMs.
Choosing either $a$-type or $c$-type transitions and driving either
$K_{c}^{''}=1\rightarrow K_{c}^{'}=0$ or $K_{a}^{''}=1\rightarrow K_{a}^{'}=0$,
correspondingly will resemble optical cycling transitions for STMs
as described above. Additionally, using quantum states such as $K_{a}^{''}=0$
or $K_{c}^{''}=0$ for $a$-type and $c$-type transitions, correspondingly,
will restrict $J$ selection rules to $\triangle J=\pm1$ only (similarly
to $\triangle N=\pm1$ transitions used for laser cooling diatomic
and linear molecules). Thus, while reduced symmetry for ARMs leads
to novel challenges in repumping vibrational levels, carefully chosen
rotational transitions can significantly limit (or even eliminate)
rotational branching. As a first step in achieving photon cycling
with ARMs, it is necessary to perform high resolution spectroscopy
of the relevant electronic transitions and identify suitable bands
for achieving optical cycling. While we have already completed such
work for CaOCH\textsubscript{3} as described above, similar spectroscopy
on more complex CaOR radicals (e.g. CaOCH\textsubscript{2}CH\textsubscript{3})
can be performed in the future. A natural chiral candidate for laser
cooling is a chiral analog of calcium monomethoxide, CaOCHDT. A recent
theoretical and experimental work by Liu and coworkers \cite{liu2016dispersed}
on calcium ethoxide (CaOC\textsubscript{2}H\textsubscript{5}) confirms
that Ca-O stretching mode is the dominant vibrational loss channel
for this molecule and indicates its potential suitability for optical
cycling and laser cooling using a scheme outlined here for ARMs. 

\section{conclusions and future directions}

While the high density of vibronic states for complex polyatomic molecules
can potentially inhibit photon cycling using the $\tilde{X}-\tilde{B}$
electronic transitions, our measurements provide a strong case that
using the lowest excited electronic level ($\tilde{A}$) will enable
photon cycling and laser cooling despite the presence of spin-orbit
and Jahn-Teller interactions. In addition to precise measurements
of the vibrational branching ratios for both $\tilde{X}-\tilde{A}$
and $\tilde{X}-\tilde{B}$ electronic excitations, we have also performed
theoretical calculations that agree quite well for the $\tilde{X}-\tilde{A}$
band, where harmonic oscillator approximation remains valid for calculating
the multidimensional Franck-Condon factors. Using the analytical integral
expressions presented by Sharp and Rosenstock \cite{sharp1964franck},
we have calculated the relevant FCFs using previously available experimental
data for geometries, vibrational frequencies, and corresponding force
fields. The presence of strong optically accessible electronic transitions,
as well as ability to scatter multiple photons from a single molecule,
opens the possibility for internal state molecular manipulation and
control of the molecular motion in the laboratory frame. While radiative
force slowing and cooling requires multiple thousands of repeated
absorption-emission cycles \cite{mccarron2018laser}, Sisyphus cooling
processes as well stimulated emission slowing methods can provide
significant gains even with a limited number of optical cycles \cite{metcalf2017colloquium,metcalf2001strong}.
Stimulated optical forces arising from polychromatic light beams \cite{kozyryev2018coherent,galica2018deflection}
might be especially effective for manipulating MOR molecules with
reduced symmetries and increased complexity where scattering tens
of thousands of photons would require a large number of vibrational
repumping lasers. Additionally, rapid optical cycling of even tens
of photons can significantly improve the rate of opto-electrical cooling
methods for STMs and ARMs, which have so far relied on using vibrational
transitions with long lifetimes and therefore low scattering rates
\cite{zeppenfeld2012sisyphus,prehn2016optoelectrical}. 

Laser cooled polyatomic molecules provide an ideal starting point
for producing ultracold fundamental radicals of chemical interest
like OH, CH\textsubscript{3} and OCH\textsubscript{3}. While a dedicated
theoretical exploration is necessary for identifying the exact route
for coherent photodissociation of polyatomic molecules into underlying
constituents with minimal energy released, this approach seems feasible
with the use of the reverse of STIRAP, which is the process employed
for forming ultracold diatomic molecules from ultracold alkali atoms
\cite{STIRAP2017stimulated}. An efficient pathway for SrOH zero-kinetic-energy
dissociation is currently being developed using ab initio molecular
potentials \cite{klos2018chemistry}. Moreover, modern theoretical
approaches to molecular quantum scattering calculations are reaching
the regime of larger polyatomic molecules \cite{krems2018time} and,
therefore, experimental results on collisions would play a crucial
role in further developing the field.

Furthermore, our measurements indicate the possibility of multiple
photon cycling for other polyatomic species including polyatomic molecular
ions. Internal state cooling and control of polyatomic molecular ions
can significantly benefit from laser-induced optical cycling. Polyatomic
molecular ions ScOH\textsuperscript{+}, YOH\textsuperscript{+},
as well as ScOCH\textsubscript{3}\textsuperscript{+} and YOCH\textsubscript{3}\textsuperscript{+},
are isoelectronic to the corresponding calcium and strontium compounds
\cite{YCH31991study,Yttrium1987gas} and should be highly suitable
for direct optical manipulation and internal state cooling using methods
previously demonstrated with diatomic ions \cite{odom2014broadband}.
A few particularly interesting candidates to consider are AcOH\textsuperscript{+}
and AcOCH\textsubscript{3}\textsuperscript{+}, which should be similar
in electronic structure to RaOH and RaOCH\textsubscript{3} and are
sensitive to new physics beyond the Standard Model \cite{isaev2015polyatomic},
but with a significantly longer half-life of the actinium-227 nucleus
compared to radium-227 \cite{flambaum2014strongly}.

Another interesting class of molecules to explore is mixed hypermettalic
neutral and ionic oxides of the MOM' type, where M and M' refer to
two different metal elements. Recently, quantum state controlled synthesis
of BaOCa\textsuperscript{+} has been demonstrated \cite{hudson2017synthesis}
and theoretical work on neutral MOM (same metal) molecules has been
performed motivated by the potential to search for fine structure
constant $\alpha$ variation \cite{SrOSr2013predicted}. Neutral polyatomic
molecules functionalized with optical cycling centers like Ca or Sr
could also enable interesting applications in quantum sciences \cite{kozyryev2017precision}. 

\section*{acknowledgments}

The work at Harvard has been supported by the grants from AFOSR and
NSF while the work at ASU has been supported by the HSF grant \#2018-0681.
We would like to thank L. Baum and B. Augenbraun for insightful discussions.
While preparing this manuscript we have learned of another recent
work on dispersed LIF of CaOCH\textsubscript{3} \cite{jinjun2018ISMS}.
However, the details of the measurement method, analysis and resulting
conclusions are vastly distinct. 

\section*{appendix}

\subsection*{Optimized geometry}

In Tables \ref{fig:Optimized-geometry-for-CAOH} and \ref{fig:Optimized-geometry-for-CaOCH3}
we provide a comparison between the measured and calculated geometries
for CaOH and CaOCH\textsubscript{3} in the ground and excited electronic
states. 

\begin{figure}
\centering{}%
\begin{tabular}{cccccc}
\hline 
State  & Coordinate  & TI  & TD  & Measured \cite{li1996dye}  & Ref. \cite{isaev2015polyatomic}\tabularnewline
\hline 
$\tilde{X}$  & Ca-O  & 1.9698  & 1.9656  & 1.9746  & 2.0038\tabularnewline
$\tilde{X}$  & O-H  & 0.9644  & 0.96460  & 0.9562  & 0.9333\tabularnewline
$\tilde{X}$  & $\angle$Ca-O-H  & 180.00  & 180.00  & 180.00  & 179.97\tabularnewline
$\tilde{A}$  & Ca-O  & 1.9698  & 1.9652  & 1.9532  & 1.9769\tabularnewline
$\tilde{A}$  & O-H  & 0.9648  & 0.9646  & 0.9572  & 0.9332\tabularnewline
$\tilde{A}$  & $\angle$Ca-O-H  & 180.00  & 180.00  & 180.00  & 179.97\tabularnewline
\hline 
\end{tabular}\caption{\label{fig:Optimized-geometry-for-CAOH}Optimized geometry for CaOH.}
\end{figure}
\begin{figure}
\begin{centering}
\begin{tabular}{ccccc}
\hline 
State  & Coordinate  & TI  & TD  & Measured \cite{crozet2005geometry}\tabularnewline
\hline 
$\tilde{X}$  & Ca-O  & 1.9726  & 1.9636  & $1.962\pm0.004$\tabularnewline
$\tilde{X}$  & O-C  & 1.3989  & 1.4011  & $1.411\pm0.007$\tabularnewline
$\tilde{X}$  & C-H  & 1.1069  & 1.1062  & $1.0937^{\dagger}$\tabularnewline
$\tilde{X}$  & $\angle$O-C-H  & 111.545  & 111.3  & $111.3\pm0.2$\tabularnewline
$\tilde{A}$  & Ca-O  & 1.97275  & 1.9636  & 1.94193\tabularnewline
$\tilde{A}$  & O-C  & 1.3989  & 1.401  & 1.4106\tabularnewline
$\tilde{A}$  & C-H  & 1.1068  & 1.1061  & 1.0923\tabularnewline
$\tilde{A}$  & $\angle$O-C-H  & 111.545  & 111.3  & 111.0004$^{\ddagger}$\tabularnewline
\hline 
\end{tabular}
\par\end{centering}
\caption{\label{fig:Optimized-geometry-for-CaOCH3}Optimized geometry for CaOCH\protect\protect\protect\textsubscript{3}.
$^{\dagger}$Using methanol tabulation. $^{\ddagger}$Calculated from
the available data for the H-C-H angle. }
\end{figure}

\subsection*{Details of FCF calculations}

To compute the FCFs for CaOH and CaOCH$_{3}$, we used Wilson's GF
matrix method and the Sharp-Rosenstock expansion of the FC overlap
integral using a harmonic potential. In depth discussion and derivation
of these methods are extensively addressed elsewhere \cite{wilson1980molecular,sharp1964franck,weber2003franck}.
Importantly, the Sharp-Rosenstock expansion depends on the vibrational
coordinates of the initial and final states, denoted $\mathbf{Q}'$
and $\mathbf{Q}$ respectively. The relationship between these two
coordinate systems is: 
\begin{equation}
\mathbf{Q}'=\mathbf{J}\mathbf{Q}+\mathbf{K}
\end{equation}
where $\mathbf{K}$ is the difference between the equilibrium geometry
of the two states in terms of the initial vibrational coordinates,
while $\mathbf{J}$ accounts for coordinate mixing, also known as
the Duchinsky effect. 

Wilson's GF matrix method \cite{ng1994unimolecular} is the extension
of eigenvalue vibration problems to molecular vibrations, and can
be used to find the relationship between $\mathbf{Q}$ and $\mathbf{Q}'$.
For an $N$-atom molecule, we define a set of $3N-6$ internal coordinates
$\{S_{t}\}$ transformed by $3N$ coefficients $B_{ti}$ from the
set of $3N$ Cartesian components $\{\xi_{i}\}$\footnote{Three degrees of freedom are taken up for position, three for rotation.
Accordingly, a linear molecule has $3N-5$ degrees of freedom due
to rotational invariance around the primary symmetry axis.}
\begin{align}
S_{t} & =\sum_{i=1}^{3N}B_{ti}\xi_{i}\hspace{0.5cm}t=1,2,3...,3N-6.
\end{align}
For convenience, the internal coordinates are often rewritten as $N$
vectors $\mathbf{s}_{t\alpha}=(B_{t\,i},B_{t\,i+1},B_{t\,i+2})$ for
each atom given fixed $t\in[1,3N-6]$. The elements of the $\mathbf{G}$
(geometry) matrix are derived from mass-weighting the summed scalar
product of each of the $3N-6$ internal coordinates over all atoms,
while the $\mathbf{F}$ (force) matrix elements are derived using
a harmonic approximation -{}- each element of the $\mathbf{F}$ matrix
is the second derivative of the potential energy surface with respect
to the internal coordinates. 
\begin{align}
 & G_{tt'}=\sum_{i=1}^{3N}\frac{1}{m_{\alpha}}B_{ti}B_{t'i}=\sum_{\alpha=1}^{N}\frac{1}{m_{\alpha}}\mathbf{s}_{t\alpha}\cdot\mathbf{s}_{t'\alpha} & F_{tt'}=\frac{\partial^{2}V}{\partial S_{t}\partial S_{t'}}
\end{align}
From the $\mathbf{G}$ matrix, it is possible to derive the kinetic
energy, and from the $\mathbf{F}$ matrix it is possible to derive
the potential energy. 
\begin{align}
 & 2T=\sum_{tt'}(G_{tt'}^{-1})\dot{S}_{t}\dot{S}_{t'} & 2V=\sum_{tt'}F_{tt'}S_{t}S_{t'}
\end{align}
Writing the Lagrangian of the system, we end up with a determinant
problem: 
\begin{align}
|\mathbf{F}-\mathbf{G}^{-1}\lambda|=\begin{vmatrix}F_{11}-(G^{-1})_{11}\lambda & \cdots & F_{1n}-(G^{-1})_{1n}\lambda\\
F_{21}-(G^{-1})_{21}\lambda & \cdots & F_{2n}-(G^{-1})_{2n}\lambda\\
\vdots & \vdots & \vdots\\
F_{n1}-(G^{-1})_{n1}\lambda & \cdots & F_{nn}-(G^{-1})_{nn}\lambda
\end{vmatrix}=0
\end{align}
where $\lambda=4\pi^{2}\nu^{2}$ and $\nu$ is the frequency of a
molecular vibration. Multiplying by the determinant of $\mathbf{G}$,
we get a diagonalization problem: $|\mathbf{G}||\mathbf{F}-\mathbf{G}^{-1}\lambda|=|\mathbf{GF}-\lambda\mathbb{I}|=0$,
where the corresponding eigenvectors are the vibrations (normal modes)
in terms of the internal coordinates. Thus, the linear transformation
$\mathbf{L}$ formed by the eigenvectors maps from vibrational normal
mode coordinates to internal coordinates. By repeating the GF analysis
for the initial and final electronic states, we end up with $\mathbf{L}$
matrices for both states, and thus can define $\mathbf{Q}$ and $\mathbf{Q}'$. 

Specifically, $\mathbf{J}$ is defined as the product of the inverse
of the initial state $\mathbf{L}$ matrix and and the final state
$\mathbf{L}$, while $\mathbf{K}$ is the transformation of the equilibrium
difference in internal coordinates to normal coordinates. 
\begin{align}
 & \mathbf{J}=(\mathbf{L}')^{-1}\mathbf{L} & \mathbf{K}=(\mathbf{L}')^{-1}(\mathbf{R}_{\text{eq}}-\mathbf{R}_{\text{eq}}')
\end{align}
Due to vibrational degeneracies, it will often be preferable to reduce
the dimensions of the GF problem by concatenating the internal coordinates
into symmetry coordinates shown in Table \ref{tab:Symmetry-coordinates-for-MOR}. 

To summarize, these Franck-Condon calculations required knowledge
of the molecule geometry, vibrational frequencies, and the potential
energy surface in the form of harmonic force constants defined over
an internal coordinate system. Suitable internal and symmetry coordinate
systems were analytically derived using literature on methanol (HOCH$_{3}$)
\cite{zhao1995methanol}, and the force constants were fit using PGOPHER
software to molecular geometry and spectroscopic data \cite{western2017pgopher}. 

The force constant fitting was initialized by first reducing the dimensionality
of the GF problem to a triatomic situation by treating the CH$_{3}$
methyl group as a single atom. Given an ordering of the eigenvalues
and neglecting off-diagonal $\mathbf{F}$ matrix interactions, the
$3\times3$ GF eigenvalue problem is completely determined without
knowledge of the diagonal matrix terms. We determine the correct ordering
by comparing calculated FC values with known measurements and prior
theory, initializing the PGOPHER force constant fitting using force
values for the M-O stretch, O-CH$_{3}$ stretch, and M-O-C bend symmetry
coordinates. To prevent misidentification of normal modes with frequencies,
we separately fit the force constants for vibrations of different
symmetries, which should be completely decoupled in such a way that
the $\mathbf{L}$ matrix is block-diagonal. We then computed the FC
overlap integrals using the Sharp-Rosenstock expansion with the matrices
$\mathbf{J}$, $\mathbf{K}$, and $\boldsymbol{\Gamma}$, a diagonal
matrix consisting of $i$ reduced vibrational frequencies $4\pi^{2}\nu_{i}/h$,
for both the initial and final electronic states. The FCFs were derived
by calculating and normalizing the overlap integral for each vibration
up to the third quantum number.

\section*{References}

\bibliographystyle{ieeetr}
\bibliography{CaOR_library_v2.bib}

\end{document}